\documentclass[reprint, aps, prl, floatfix,nofootinbib, superscriptaddress]{revtex4-2}
 
\usepackage[T1]{fontenc}
\usepackage[utf8]{inputenc}
\usepackage{lmodern}
\usepackage{setspace}
\usepackage[marginal, multiple]{footmisc}
\usepackage[colorlinks, allcolors=blue!70!black, linktocpage]{hyperref}

\usepackage{amsmath, amssymb, amsfonts, amsthm, latexsym, mathrsfs, xcolor, bbm, slashed, braket, cancel, graphicx, mathtools, setspace, orcidlink, enumitem}

\newcommand{\be}{\begin{equation}\begin{aligned}}
\newcommand{\ee}{\end{aligned}\end{equation}}

\newcommand{\lie}[1]{\mathcal{L}_{#1}}

\usepackage[allcommands,old-arrows]{overarrows}
\newcommand{\pullback}[1]{\underleftarrow{#1}}

\usepackage{soul}
\newcommand{\comment}[1]{{#1}}

\begin{document}
\count\footins = 800

\title{On Radiative Fluxes and Coulombic Charges \\ in the Balance Law for Black Hole Evaporation}

\author{Eugenio Bianchi \orcidlink{0000-0001-7847-9929}}
\email{ebianchi@psu.edu}
\affiliation{Institute for Gravitation \& the Cosmos\\ and Department of Physics, The Pennsylvania State University\\ University Park, PA 16802, USA}

\author{Daniel E. Paraizo \orcidlink{0000-0002-8653-0971}}
\email{dparaizo@psu.edu}
\affiliation{Institute for Gravitation \& the Cosmos\\ and Department of Physics, The Pennsylvania State University\\ University Park, PA 16802, USA}
\affiliation{Applied Research Lab, The Pennsylvania State University\\University Park, PA 16802, USA}

%%%%%%%%%%%%%%%%%%%%%%%%%%%%%%%

\begin{abstract}
  In asymptotically-flat spacetimes, there is a clear distinction between radiative fluxes and Coulombic charges. Using the Wald-Zoupas prescription, we identify the classical radiative flux of a massless scalar field in 3+1 dimensions. In a spherically-symmetric model of black hole evaporation, the balance law yields a Bondi mass correction related to the entanglement entropy of Hawking radiation. The renormalized flux is nonnegative, differs from the Fulling-Davies formula, and coincides with the Ashtekar-Taveras-Varadarajan flux. We discuss implications for 3+1 black hole evaporation.
\end{abstract}

\maketitle

%%%%%%%%%%%%%%%%%%%%%%%%%%%%%%%

\emph{Introduction}---An evaporating black hole loses mass at a rate $\dot{M}= - F$ given by the balance law for the energy flux  $F$ carried by Hawking radiation \cite{Hawking:1974rv,Hawking:1975vcx, wald_1975, PhysRevD.14.870, Page:1976df}. In an asymptotically-flat spacetime, the mass and the radiative flux acquire a clear operational meaning when understood as physical quantities measured at future null infinity \cite{Bondi:1960jsa,Bondi:1962rkt,Sachs:1962zza}. Specifically, even though a sufficiently small black hole emits thermal electrons and positrons, this flux of massive particles cannot contribute to the radiative Bondi flux and, in fact, it corresponds to a Coulombic contribution to the Bondi mass of the black hole. 

In this Letter, we show that---even for a massless scalar field that is minimally coupled to gravity in $3+1$ dimensions---the Bondi mass of an evaporating black hole has a Coulombic correction which contributes to the balance law for the renormalized expectation value \cite{BirrellandDavies,FabbriandSalas, Wald_QFTCS, Wald_ST_Axioms, WALD1978472} of the radiative flux. We identify the classical origin of this phenomenon by investigating the \comment{Wald-Zoupas (WZ) prescription \cite{Wald_2000}} for the balance law of Bondi–Metzner–Sachs (BMS) fluxes and the associated Coulombic charges \cite{Bondi:1962px, Flanagan:2015pxa, Compere:2019gft, CompereNotes, Bonga_2020, Odak:2022ndm, Rignon-Bret:2024gcx}, using the intrinsic formulation at future null infinity \cite{Geroch1977,Ashtekar-Streubel,Ashtekar_AQ,Ashtekar:1987tt,Ashtekar:2014zsa,Ashtekar_2024}.

After clarifying the classical balance law in $3+1$ dimensions, we investigate black hole evaporation adopting the usual spherically-symmetric $s$-wave model. We show that the quantum correction to the Bondi mass is related to the entanglement entropy of the Hawking radiation studied in \cite{Bianchi:2014vea,Bianchi:2014qua,Bianchi:2014bma, Abdolrahimi_2015}. Within this model of $3+1$ black hole evaporation, we find that the renormalized expectation value of the radiative flux turns out to be manifestly positive and it takes the same form as the flux proposed by Ashtekar, Taveras, and Varadarajan for 2d dilatonic black hole evaporation \cite{Ashtekar:2008jd,Ashtekar:2010hx,Ashtekar:2010qz,Barenboim:2025fds}, thus providing a new classical $3+1$ dimensional justification for this proposal.

The spherically-symmetric reduction used in the quantum theory is equivalent to filling-in the bulk $3+1$ dimensional spacetime of an evaporating black hole via an effective semiclassical geometry given by the Fulling-Davies moving mirror model \cite{Davies-Fulling_MM, Davies-Fulling_MM+BH, Davies-Fulling76,Varadarajan-2025}, which provides a map from past to future null infinity \cite{Walker1985, Carlitz-Wiley1987_Lifetime, Carlitz-Wiley1987_Reflections, FordRoman_MM_Neg_Energy, wilczek1993, Ford_2004, Good_2013, Good_2015, Good2016, Chen_2017, Good2018, Good2019, Wald_2019} (see also \cite{Hotta_PartnerParticles,Tomitsuka_2020,Agullo_2024,Agullo:2026}). We show that the expression of our radiative flux applies beyond the adiabatic approximation, where it does not coincide with the standard Fulling-Davies formula, and discuss its consequences for the end-point of black hole evaporation. Throughout the paper we report factors of the Planck constant $\hbar$ to tell apart quantum from classical phenomena, but do not include factors of $c$, $G_N$, and $k_B$ which can be easily restored. With these conventions, the Planck mass is $m_P=\sqrt{\hbar}$.

%%%%%%%%%%%%%

\smallskip

\emph{Radiative Fluxes and Coulombic Charges}---We consider General Relativity with a minimally-coupled massless scalar field $\varphi$. The $4d$  spacetime manifold is denoted $\mathcal{M}$, and the Lorentzian metric $g_{ab}$ has signature ($-$+++). The Lagrangian 4-form is 
\begin{equation}\label{lagrangian}
    L = \big( \tfrac{1}{16 \pi} R - \tfrac{1}{2} g^{ab} \nabla_a \varphi \nabla_b \varphi \big)\, \epsilon_4 ,
\end{equation}
which determines the stress tensor $T_{ab}=\nabla_a\varphi\,\nabla_b\varphi-\tfrac{1}{2}g_{ab}\,g^{cd}\,\nabla_c\varphi\nabla_d\varphi$.
We assume that the physical spacetime $(\mathcal{M},g_{ab})$ is asymptotically flat at (past and future) null infinity $\mathcal{I}$. This condition can be formulated by introducing an auxiliary spacetime $(\tilde{\mathcal{M}},\tilde{g}_{ab})$ defined by a conformal completion (see \cite{Geroch1977,Ashtekar-Streubel,Ashtekar_AQ,Ashtekar:1987tt,Ashtekar:2014zsa,Ashtekar_2024} for reviews and definitions). As usual, one introduces a conformal factor $\Omega$ and the rescaled fields $\tilde{g}_{ab}=\Omega^2 g_{ab}$, $\tilde{\varphi}=\Omega^{-1}\varphi$ that admit finite limits to $\mathcal{I}$. Without loss of generality, we will work with a conformal completion which satisfies the divergence-free condition $\tilde{g}^{ab}\,\tilde{\nabla}_a\tilde{\nabla}_b\Omega=0$ on $\mathcal{I}^+$. Null infinity is defined as the boundary manifold  $\partial\tilde{\mathcal{M}}=\mathcal{I}$. Bulk fields on $\mathcal{M}$ induce boundary fields at future null infinity $\mathcal{I}^+$ via pullback (denoted by an under arrow, e.g., $\pullback{w_a}$ for a bulk one-form $w_a$).  The induced metric at $\mathcal{I}^+$ is denoted $q_{ab}=\pullback{\tilde{g}_{ab}}$, the induced covariant derivative $D_a \pullback{w_b}=\pullback{\tilde{\nabla}_a w_b}$, the null normal $n^a=\pullback{\,\tilde{g}^{ab}\,\tilde{\nabla}_b\Omega}$, and the scalar radiation field $\chi=\pullback{\tilde{\varphi}}$.

\smallskip

The relevant structures that determine the radiative flux of the scalar field can be defined intrinsically on the $3d$ manifold $\mathcal{I}^+$, which has topology $S^2\times \mathbb{R}$ \cite{Geroch1977,Ashtekar-Streubel,Ashtekar_AQ,Ashtekar:1987tt,Ashtekar:2014zsa,Ashtekar_2024}. The intrinsic metric $q_{ab}$ has signature (0\,++), and the normal $n^a$ satisfies the condition $q_{ab}n^b=0$. The metric does not fully determine the torsion-free covariant derivative $D_a$, which encodes the radiative degrees of freedom of the gravitational field. This covariant derivative operator covariantly conserves the structures $(q_{ab},n^a,\epsilon^{abc})$, i.e., $D_c\, q_{ab}=0$, $D_a n^b=0$, $D_a\,\epsilon^{mnp}=0$, where $\epsilon_{abc}$ is (up to a sign)  the unique volume $3$-form at $\mathcal{I}^+$ satisfying $\epsilon^{abc}\epsilon^{mnp}q_{am}q_{bn}=n^c n^p$ with normalization $\epsilon_{abc}\epsilon^{abc}=6$. Note that there is a freedom in the choice of the conformal factor $\Omega$ that defines the auxiliary spacetime. In particular, a conformal rescaling $\Omega \rightarrow \omega \,\Omega$ with $\omega$ smooth and non-zero on $\mathcal{I}^+$ defines the same asymptotically flat spacetime $(\mathcal{M},g_{ab})$. The divergence-free condition restricts the allowed conformal rescalings to a residual conformal freedom $\omega$ satisfying $n^a D_a\omega=0$. Under these rescalings, the intrinsic structures at $\mathcal{I}^+$ transform as
\begin{align}
\label{eq:rescaling}
q_{ab}&\,\to \omega^2\, q_{ab}\,,\quad n^a\to \omega^{-1}\,n^a\,,\quad \epsilon_{abc}\to \omega^3\,\epsilon_{abc}\,,\\[.2em] 
\chi&\,\to \omega^{-1}\,\chi\,,\quad \textrm{with}\quad n^a D_a\omega=0\,.\nonumber
\end{align}
All radiative observables are required to be invariant under the residual conformal transformations \eqref{eq:rescaling}. The infinitesimal version of these transformations determines the BMS vector field $\xi^a$, defined intrinsically at $\mathcal{I}^+$ \cite{Geroch1977,Ashtekar-Streubel,Ashtekar_AQ,Ashtekar:1987tt,Ashtekar:2014zsa,Ashtekar_2024}. Specifically, one has that the Lie derivatives in the direction $\xi^a$ are $\mathcal{L}_\xi q_{ab}=2 \alpha\, q_{ab}$, $\mathcal{L}_\xi n^a=-\alpha\, n^a$, and $\mathcal{L}_\xi \epsilon^{abc}=-3 \alpha\, \epsilon^{abc}$, with $n^a D_a \alpha=0$. Note that, as $\mathcal{L}_\xi \epsilon^{abc} =-(D_m\xi^m)\,\epsilon^{abc}$ and $\mathcal{L}_\xi n^a=-n^bD_b\xi^a$, we have also the relation $\alpha=\frac{1}{3}D_a\xi^a$.

\smallskip

\comment{To motivate our analysis, note that expressing the stress tensor $T_{ab}$ in terms of the rescaled fields and pulling back to $\mathcal{I}^+$ will determine a stress-tensor flux $\mathcal{F}^{(\mathrm{st}
)}_{\mathrm{mat}}[\Delta\mathcal{I},\xi]$ through a portion $\Delta\mathcal{I}$ of future null infinity. Specifically, the stress-tensor flux in the direction of the BMS vector field $\xi^a$ is given by
\begin{equation}
\label{eq:Fmatt}
\mathcal{F}^{(\mathrm{st}
)}_{\mathrm{mat}}[\Delta\mathcal{I},\xi]=\!\int_{\Delta\mathcal{I}}(D_a\chi)(D_b\chi)\,n^a\,\xi^b\,\,\epsilon_{mnp}\,d\mathcal{I}^{mnp}\,.
\end{equation}
However, this flux is \textit{not} purely radiative as it is not invariant under the residual conformal rescalings \eqref{eq:rescaling} for a general BMS vector field. In particular, note that while Eq.~\eqref{eq:Fmatt} is conformally invariant for a supertranslation, the boost-momentum is not. This means the \textit{same physical spacetime would admit different fluxes for the same field configuration}; thus, the stress-tensor flux 
$\mathcal{F}^{(\mathrm{st}
)}_{\mathrm{mat}}$, Eq.~\eqref{eq:Fmatt}, is not a viable definition of the radiative flux at future null infinity.

Fortunately, there is a well-developed procedure to obtain fluxes from the symplectic structure of the covariant phase space using Noether's theorem, originally given by Wald-Zoupas (WZ) \cite{Wald_2000} and expanded on by many authors \cite{Harlow:2019yfa, Bonga_2020, Grant:2021sxk, Odak:2022ndm, Odak:2023pga, Rignon-Bret:2024gcx, Rignon-Bret-Speziale2025}. We refer to the review \cite{Speziale:2025lkm} for definitions and conventions. We take the symplectic structure for the gravity sector to be the standard one, see e.g. \cite{Wald_2000}. Focusing on the matter contribution, we denote the \textit{reference} symplectic potential (i.e., the one obtained directly from varying the Lagrangian \eqref{lagrangian}), pulled back to $\mathcal{I}^+$, as $\pullback{\theta_{\mathrm{mat}}}(\delta \chi)=-(n^a D_a \chi)\,\delta \chi\, \epsilon_3$, with $\epsilon_3=\epsilon_{mnp}d\mathcal{I}^{mnp}$. Crucially, however, the Lagrangian does not fully determine the symplectic structure of the theory. Indeed, one is free to add a boundary term $d  \ell$ to the Lagrangian, and a corner term $d \vartheta$ to the reference potential; such terms would change the potential, but not the equations of motion \cite{Speziale:2025lkm}. Hence, one can always choose a \textit{preferred} symplectic potential $\Theta$ that differs from the reference potential by these ambiguities: $\pullback{\theta} = \Theta - \delta \ell + d \vartheta$. The generalized WZ prescription gives a set of admissibility criteria to choose this preferred potential. Namely, $\Theta$ must satisfy the following four conditions: (i) be a potential for the pullback of the symplectic current, i.e., $\pullback{\omega}(\delta_1 \chi, \delta_2 \chi) = \delta_1 \Theta(\delta_2 \chi) - \delta_2 \Theta(\delta_1 \chi)$; (ii) be locally and covariantly constructed from the dynamical fields and their first-order perturbations; (iii) be independent from arbitrary choices of background structure and, specifically, it must be invariant under the rescalings \eqref{eq:rescaling}; (iv) vanish for stationary solutions $n^aD_a \chi =0$ when evaluated on $\delta_\xi \chi$, i.e., a BMS variation. Once this potential is determined, the WZ flux is
\begin{equation}\label{WZ-flux}
    \mathcal{F}[\Delta \mathcal{I}, \xi] = -\int_{\Delta \mathcal{I}} \Big( \Theta_{\mathrm{grav}}(\delta_\xi g) +\Theta_{\mathrm{mat}}(\delta_\xi \chi) \Big)\,.
\end{equation}
The Einstein equations $G_{ab}=8\pi\, T_{ab}$ at future null infinity then take the form of a balance law
\begin{equation}
\label{WZ-balance}
    Q[S_2,\xi]-Q[S_1,\xi]=-\mathcal{F}[\Delta \mathcal{I}, \xi],
\end{equation}
where $S_1$ and $S_2$ are cross-sections on the boundary of $\Delta\mathcal{I}$, i.e., $(-S_1)\cup S_2=\partial(\Delta\mathcal{I})$, and the charge on a cross-section $S$ is given by the sum of the gravitational and the matter charge:
\begin{equation}
    Q[S, \xi]=Q_{\rm grav}[S, \xi] + Q_{\rm mat}[S, \xi].
\end{equation}
The conditions (i)--(iv) do not yet uniquely determine the potential for the scalar field, as they admit the one-parameter family
\begin{equation}\label{pref_pot}
    \Theta_{\mathrm{mat}}(\delta \chi) = - \big(  (n^a D_a \chi)\delta \chi + \beta\, \delta (\chi\, n^a D_a \chi) \big) \,\epsilon_3,
\end{equation}
where $\beta$ is a real number. By inspection, the one-parameter ambiguity is due to a boundary term $\delta(\lie{n} \chi^2)$, or alternatively, a $\delta$-exact corner term. Previously, this ambiguity was fixed by imposing a stronger version of condition (iv). Namely, (iv') the potential should vanish on stationary solutions for \textit{all} perturbations $\delta \chi$, which would then pick-out $\beta = 0$, \cite{Wald_2000}. However, the flux would then depend on derivatives of $\xi$ since $\delta_\xi \chi = \xi^aD_a \chi + \frac{1}{3}\, \chi D_a \xi^a$. Instead, since we are interested in the local flux in one direction, we impose the following additional condition (v): the evaluated potential $\Theta_{\mathrm{mat}}(\delta_\xi \chi)$ must be ultra-local in $\xi$, i.e., have no derivatives of the BMS direction $\xi$, so that the flux has the form $\mathcal{F}^{(\mathrm{rad})}_{\mathrm{mat}} = \int j_a\, \xi^a \epsilon_3$. Within the family \eqref{pref_pot}, this condition uniquely fixes the value $\beta = -1/3$.

Combining Eq.~\eqref{WZ-flux} and \eqref{pref_pot}, together with condition (v), we find that the radiative flux of the scalar field is
\begin{align}
\label{eq:F-Rad}
&\mathcal{F}^{(\mathrm{rad})}_{\mathrm{mat}}[\Delta\mathcal{I},\xi] \equiv -\int_{\Delta \mathcal{I}} \Theta_{\rm mat}(\delta_\xi \chi) \\[.2em]
&=\!\int_{\Delta\mathcal{I}}\!\Big(\tfrac{2}{3}(D_a\chi)(D_b\chi)-\tfrac{1}{3}(\chi D_aD_b\chi)\Big)
 n^a\,\xi^b\,\,\epsilon_{mnp}\,d\mathcal{I}^{mnp}.\nonumber
\end{align}
The associated Coulombic charge on a cross-section $S$ is 
\begin{equation}
\label{eq:Q-phi}
Q_{\mathrm{mat}}[S,\xi]=\int_S (\chi D_a \chi)(\tfrac{1}{2}n^b\,\xi^a-\tfrac{1}{6}n^a\,\xi^b)\,\epsilon_{bmn}\,dS^{mn}.
\end{equation}
This is the first main result of this paper. Note that the stress-tensor flux \eqref{eq:Fmatt} differs from the radiative flux \eqref{eq:F-Rad}  precisely by the boundary term $Q_{\mathrm{mat}}[S_2,\xi]-Q_{\mathrm{mat}}[S_1,\xi]$.
}

%We refer to equations (4.26)--(4.27) in \cite{Bonga_2020} for precise expressions of $\mathcal{F}^{(\mathrm{rad})}_{\mathrm{grav}}$ and $Q_{\mathrm{grav}}[S,\xi]$; they will not be needed in full generality. Therefore, as required, the full WZ flux-formula \eqref{WZ-flux} receives no contribution from the stress tensor, and can be written purely as a difference of (gravitational plus matter) charges, i.e. a balance law.

\smallskip

In calculations, it is useful to introduce Bondi-Sachs coordinates $x^\mu=(u,r,\theta,\phi)$, with $u$ the retarded time at future null infinity, $\theta^A=(\theta,\phi)$ angular coordinates on the unit sphere with metric $\gamma_{AB}$, and $r$ the areal radius \cite{Bondi:1962px,Flanagan:2015pxa, Compere:2019gft, CompereNotes}. As usual, the metric takes the asymptotic form $g_{\mu\nu}dx^\mu dx^\nu=-\big(1-\frac{2m(u,\theta,\phi)}{r}+\ldots\big)du^2-2(1+\ldots)dudr+
r^2\big(\gamma_{AB}(\theta,\phi)+\frac{C_{AB}(u,\theta,\phi)}{r}+\ldots)d\theta^Ad\theta^B+(D_AC^A{}_B+\ldots\,)dud\theta^B$, where $m(u,\theta,\phi)$ is the mass aspect, $C_{AB}(u,\theta,\phi)$ the shear tensor, and $N_{AB}\equiv \partial_u C_{AB}(u,\theta,\phi)$ the Bondi news tensor. Similarly, the scalar field is assumed to have the asymptotic form
\begin{equation}
\varphi(u,r,\theta,\phi)\;=\;\frac{\chi(u,\theta,\phi)}{r}\;+\;\ldots\,,
\end{equation}
where the dots $\ldots$ stand for terms that fall off more rapidly as $r\to\infty$. As we are interested in the energy flux, we restrict attention to the BMS vector field for a translation along the null direction $n^a$, i.e., $\xi^a=n^a=(\partial_u)^a$. For a cross-section $S$ of constant retarded time $u$, the radiative fluxes for gravity and for matter are
\begin{align}
\label{eq:F-rad-grav}
F^{(\mathrm{rad})}_{\mathrm{grav}}(u)=&\int\tfrac{1}{32\pi}(\partial_u C^{AB})(\partial_u C_{AB})\,\sin\theta\,d\theta d\phi\,,\\
\label{eq:F-rad-mat}
F^{(\mathrm{rad})}_{\mathrm{mat}}(u)=&\int\!\Big(\tfrac{2}{3}\partial_u \chi\,\partial_u\chi-\tfrac{1}{3}\chi \partial_u\partial_u \chi\Big)\,\sin\theta\,d\theta d\phi\,,
\end{align}
where the first expression is the familiar Bondi flux of gravitational waves \cite{Bondi:1962px,Flanagan:2015pxa, Compere:2019gft, CompereNotes} and the second is the radiative flux for the scalar field determined above, Eq.~\eqref{eq:F-Rad}. The WZ flux formula and balance law \eqref{WZ-flux}--\eqref{WZ-balance} determine the mass-loss formula
\begin{equation}
\label{eq:M-dot}
\partial_u M(u)= -\mathcal{F}^{(\mathrm{rad})}_{\mathrm{grav}}(u)- \mathcal{F}^{(\mathrm{rad})}_{\mathrm{mat}}(u)\,,
\end{equation}
where the Bondi mass $M(u)$ is defined by the balance law for the radiative fluxes of gravity and matter appearing on the right hand side. \comment{Surprisingly, we find that the Bondi mass $M(u)$ is not purely determined by the mass aspect $m(u,\theta,\phi)$, but it includes a contribution from the Coulombic charge $Q_{\mathrm{mat}}[S,\xi]$ of the scalar field, Eq.~\eqref{eq:Q-phi}:}
\begin{equation}
\label{eq:M-Bondi}
M(u)=\!\int\!\!\Big(\frac{m(u,\theta,\phi)}{4\pi}+\tfrac{1}{3}\chi\partial_u\chi(u,\theta,\phi)\Big)\sin\theta\,d\theta d\phi.
\end{equation}
This result in the classical theory, based on the notion that radiative fluxes are more fundamental than charges \cite{Ashtekar_2024}, plays a central role in the balance law for black hole evaporation which we discuss next.

\smallskip

\emph{Radiative Fluxes in Black Hole Evaporation}---In the quantum theory, the emission of Hawking radiation results in fluxes of energy that change the mass of the black hole. The distinction between radiative fluxes and Coulombic charges gives a precise physical characterization of what one means by mass and energy flux: A radiation detector placed far away from an evaporating black hole will only measure the radiative fluxes $F^{(\mathrm{rad})}_{\mathrm{grav}}(u)$ and $F^{(\mathrm{rad})}_{\mathrm{mat}}(u)$. The changes in the Bondi mass $M(u)$ can then be \textit{inferred} from the energy carried away by these fluxes.

As a first step, one needs the two-point correlation functions for the perturbative quantum fields $N_{AB}(u,\theta,\phi)$ and $\chi(u,\theta,\phi)$ in a state $|\psi\rangle$ in the asymptotic Hilbert space of the radiative degrees of freedom \cite{Ashtekar:1987tt, Ashtekar_AQ}. Given this, one can compute the renormalized expectation values  $\langle F^{(\mathrm{rad})}_{\mathrm{grav}}\rangle(u)$ and $\langle F^{(\mathrm{rad})}_{\mathrm{mat}}\rangle(u)$ of the radiative fluxes \eqref{eq:F-rad-grav}--\eqref{eq:F-rad-mat} at future null infinity. The mass-loss formula is then given by the quantization of the classical expression \eqref{eq:M-dot}:
\begin{equation}
\label{eq:Mu-MADM}
\langle M\rangle(u) = M_{\mathrm{ADM}}-\!\int_{-\infty}^{u}\!\!\langle F^{(\mathrm{rad})}_{\mathrm{grav}}+ F^{(\mathrm{rad})}_{\mathrm{mat}}\rangle(u')\,du'\,,
\end{equation}
where $M_{\mathrm{ADM}}$ is the ADM mass and $\langle M\rangle(u)$ is the renormalized expectation value of the Bondi mass \eqref{eq:M-Bondi}.

In black hole evaporation, one assumes the perturbative quantum fields to be  prepared in the in-vacuum $|0_{\mathrm{in}}\rangle$. The dynamics provides a map from $\mathcal{I}^-$ to $\mathcal{I}^+$ which determines the correlation functions {\small $\langle 0_{\mathrm{in}}|N_{AB}(u,\theta,\phi)N_{CD}(u',\theta',\phi')|0_{\mathrm{in}}\rangle$} and {\small $\langle0_{\mathrm{in}}|\chi(u,\theta,\phi)\chi(u',\theta',\phi')|0_{\mathrm{in}}\rangle$} in the future, and therefore the renormalized expectation value of the radiative fluxes. A well-studied simplified model where the correlation functions can be computed in closed form, while still capturing key features of black hole evaporation, is the mirror model \cite{Davies-Fulling_MM, Davies-Fulling_MM+BH, Davies-Fulling76,Varadarajan-2025, BirrellandDavies}.

In the mirror model for $4$d black hole evaporation, both the background classical fields and the quantum perturbations are assumed to be spherically symmetric. As a result, the flux of gravitational radiation is identically zero. For a minimally-coupled massless scalar field, one considers only $s$-wave perturbations, i.e., one restricts to the $\ell=0$ component $\chi_0(u)$ in the spherical harmonics expansion $\chi(u,\theta,\phi)=\sum_{\ell=0}^{\infty}\sum_m\chi_{\ell m}(u)\,\sqrt{4\pi} Y_{\ell m}(\theta,\phi)$. Furthermore, one neglects the backscattering potential for the perturbation $\chi_0(u)$. As a result, the geometric optics approximation is assumed to apply:  the ray-tracing function $v=p(u)$ describes how a radial light ray incoming from the advanced time $v$ at $\mathcal{I}^-$ emerges at $\mathcal{I}^+$ at the retarded time $u$. In double-null coordinates $(u,v,\theta,\phi)$, the ray-tracing function describes the axis of spherical symmetry $r(u,v)=0$. Equivalently, \comment{were one to move off $\mathcal{I}$ and `fill-in' the bulk,} the function $v=p(u)$ can be understood as the trajectory of a moving mirror in a $2$d Minkowski spacetime. The redshift factor for null rays reflecting off the axis of symmetry is given by the derivative $\dot{p}(u)\equiv\partial_u p(u)$. Its logarithmic derivative, 
\begin{equation}
\label{eq:ku}
k(u)\equiv -\ddot{p}(u)/\dot{p}(u)\,,
\end{equation}
defines the peeling function $k(u)$ \cite{Barcel2011-HawkRad, Barcel2011-Minimal_Conditions}.

For this model, the momentum two-point correlation function of the scalar field at $\mathcal{I}^+$ is given by $\langle 0_\mathrm{in}|\partial_u\chi_0(u)\,\partial_{u'}\chi_0(u')|0_\mathrm{in}\rangle=-\frac{1}{4\pi}\frac{\hbar}{4\pi}\frac{\dot{p}(u)\dot{p}(u')}{(p(u)-p(u'))^2}$ \cite{ wilczek1993, Holzhey_1994}. The renormalized expectation value of quadratic operators can be obtained from this correlation function via point-splitting in $u$ and subtraction with respect to the out-vacuum $|0_{\mathrm{out}}\rangle$. In particular, assuming that $\dot{p}(u)\to 1$ for $u\to-\infty$, one finds the renormalized expectation values at $u$ to be:
\begin{align}
\langle \partial_u\chi_0\,\partial_u\chi_0\rangle(u)&\,=\tfrac{1}{4\pi}\;\tfrac{\hbar}{48\pi}\big(k^2(u)+2\dot{k}(u)\big)\,,\label{eq:DxDx}\\[.3em]
\langle \chi_0\,\partial_u\partial_u\chi_0\rangle(u)&\,=\tfrac{1}{4\pi}\;\tfrac{\hbar}{48\pi}\big(\!-k^2(u)+4\dot{k}(u)\big)\,,\label{eq:xDDx}\\[.3em]
\langle \chi_0\,\partial_u\chi_0\rangle(u)&\,=\tfrac{1}{4\pi}\;\tfrac{\hbar}{8\pi}k(u)\,.\label{eq:xDx}
\end{align}
As a result, the renormalized expectation value of the radiative flux \eqref{eq:F-rad-mat} is simply given by
\begin{align}
\label{eq:F-rad-quantum}
\langle F^{(\mathrm{rad})}_{\mathrm{mat}}\rangle(u)&=
4\pi\,\big(\tfrac{2}{3}\langle\partial_u \chi_0\,\partial_u\chi_0\rangle-\tfrac{1}{3}\langle\chi_0 \partial_u\partial_u \chi_0\rangle\big)\\[.5em]
&=\tfrac{\hbar}{48\pi}k^2(u)\,,\nonumber
\end{align}
where there is a surprising cancellation of the $\dot{k}(u)$ terms appearing in \eqref{eq:DxDx} and \eqref{eq:xDDx}. We note that the formula \eqref{eq:F-rad-quantum} for the radiative energy flux of Hawking radiation is new and does not coincide with the well-known Fulling-Davies formula \cite{Davies-Fulling76, Davies-Fulling_MM, Davies-Fulling_MM+BH}. 

The requirement that the flux is purely radiative results in the mass-loss formula
\begin{equation}
\label{eq:M-dot-quantum}
\partial_u \langle M\rangle(u)= -\tfrac{\hbar}{48\pi}k^2(u)\,.
\end{equation}
The expectation value of the semiclassical Bondi mass is \emph{defined by} the balance law, with radiative fluxes understood as fundamental and Coulombic charges as derived quantities in the classical as well in the quantum theory. Because of this, the Bondi mass receives a quantum correction from the renormalized expectation value of the Coulombic charge of the scalar field:
\begin{align}
\label{eq:M-quantum}
\langle M\rangle(u)&=m_0(u)+4\pi\,\tfrac{1}{3}\langle\chi_0\partial_u\chi_0\rangle(u)\\[.5em]
&=m_0(u)+\tfrac{\hbar}{24\pi}k(u)\,,\nonumber
\end{align}
where $m_0(u)$ is the mass aspect of the background spacetime which is assumed to be spherically symmetric, i.e., $m(u,\theta,\phi)=m_0(u)$. 

Remarkably, within the same model of black hole evaporation discussed above, one can also compute the renormalized entanglement entropy of the scalar-field radiation between the portions $(-\infty,u)$ and $(u,+\infty)$ of $\mathcal{I}^+$, which is given by the expression \cite{Bianchi:2014qua,Bianchi:2014vea,Bianchi:2014bma}
\begin{equation}
\label{eq:S-ent}
S_{\mathrm{ent}}^{(\mathrm{rad})}(u)=\int_{-\infty}^u \tfrac{1}{12} k(u')\,du'\,.
\end{equation}
Therefore, the renormalized expectation value of the Bondi mass can be equivalently written as 
\begin{equation}
\label{eq:M-entropic}
\langle M\rangle(u)=m_0(u)+\tfrac{\hbar}{2\pi}\dot{S}_{\mathrm{ent}}^{(\mathrm{rad})}(u)\,.
\end{equation}
This expression can be understood as the Bondi mass of an evaporating black hole receiving an entropic correction $\tfrac{\hbar}{2\pi}\dot{S}_{\mathrm{ent}}^{(\mathrm{rad})}(u)$ to the bare mass aspect $m_0(u)$. \comment{Intriguingly, this expression bears a striking resemblance to the Quantum Null Energy Condition at null infinity \cite{Bousso:2016vlt}, even though it is based on different considerations than the classical expression \eqref{eq:M-Bondi} for the Bondi mass.}

\smallskip

%%%% bullet version of the discussion

\emph{Discussion}---In this Letter, we introduced a new construction for radiative fluxes and we demonstrated how the balance laws determine a semiclassical notion of Bondi mass in black hole evaporation. We briefly summarize the results here:

   $\scriptstyle\bullet$ It is only with a careful analysis that properly identifies the Coulombic and radiative terms in the balance law that one can then assign a physical interpretation to what is to be understood as the mass of an evaporating black hole and the energy carried to infinity by Hawking radiation. To illustrate this, we showed that the flux defined by the stress tensor of a minimally-coupled massless scalar field includes both a radiative flux \eqref{eq:F-Rad} and a change in Coulombic charge \eqref{eq:Q-phi}, which is to be understood as a contribution to the Bondi mass of the system \eqref{eq:M-Bondi}. The mass of the black hole is inferred from balance laws \eqref{eq:M-dot} which follow from Einstein equations. 
\comment{
On the other hand, a localized detector which is switched on and off smoothly, such as a radiometer, cannot distinguish the stress-tensor flux \eqref{eq:Fmatt} from the radiative flux \eqref{eq:F-Rad} as its response function is insensitive to boundary contributions.}

    $\scriptstyle\bullet$ The quantization of the classical radiative flux \eqref{eq:F-rad-mat} defines formally the balance law for the quantum theory \eqref{eq:Mu-MADM}. To compute concretely the renormalized expectation value of the radiative flux \eqref{eq:F-rad-quantum}, we adopted the standard approximations used in the mirror model for $3+1$ black hole evaporation \cite{Davies-Fulling_MM, Davies-Fulling_MM+BH, Davies-Fulling76, Varadarajan-2025}. In this analysis, while the correlation functions are computed using a spherically-symmetric reduction to $1+1$ dimensions, the notion of radiative flux is genuinely $3+1$ dimensional. As a result, the expression we find for the radiative flux in Hawking evaporation does not coincide with the one adopted in the literature on $1+1$ models of black hole evaporation \cite{BirrellandDavies,FabbriandSalas}.

    $\scriptstyle\bullet$ The balance law \eqref{eq:M-dot-quantum} then implies that the expectation value of the Bondi mass acquires a quantum correction \eqref{eq:M-quantum} due to the Coulombic charge. Within the black-hole evaporation model considered here, this is shown to be an entropic correction \eqref{eq:M-entropic} coming from the entanglement entropy of Hawking radiation.

\smallskip

Our formula \eqref{eq:F-rad-quantum} is related to other expressions that have appeared in the literature on the energy flux of Hawking radiation. We briefly comment on the relation:

   $\diamond$ The well-known Fulling-Davies formula \cite{Davies-Fulling76,Davies-Fulling_MM, Davies-Fulling_MM+BH}, $F_{\mathrm{FD}}(u)=4\pi\,\langle\partial_u \chi_0\,\partial_u\chi_0\rangle=\tfrac{\hbar}{48\pi}\big(k^2(u)+2\dot{k}(u)\big)$ computes the energy flux due to a mirror moving in  $1+1$ dimensions. It can be understood as the quantization of the spherically-symmetric reduction of the stress-tensor flux \eqref{eq:Fmatt}. As a result, the Fulling-Davies formula includes both the radiative flux \eqref{eq:F-Rad} and the Coulombic charge \eqref{eq:Q-phi}. The two cannot be properly distinguished by working purely in $1+1$ dimensions without the structure $\mathcal{I}^+=S^2\times \mathbb{R}$ of future null infinity. 
   
   %% <<Commented for length>>
   %To see how the two notions compare, note that in the adiabatic regime where  $|\dot{k}|\ll k^2$, the peeling function $k(u)$ can be estimated by the surface gravity, i.e., $k\sim 1/M$ \cite{Barcel2011-HawkRad, Barcel2011-Minimal_Conditions, Agullo_2024, Agullo:2026}. As a result, at leading order, the flux $F\sim \hbar/M^2$ and the balance law $\dot{M}= -F$ imply $\dot{k}\sim \hbar/M^4$. Therefore the fractional difference $|F_{\mathrm{FD}}-\langle F^{(\mathrm{rad})}_{\mathrm{mat}}\rangle|/\langle F^{(\mathrm{rad})}_{\mathrm{mat}}\rangle\sim (m_P/M(u))^2$ is negligible for a large black hole, but of order one for a black hole of a few Planck masses $m_P=\sqrt{\hbar}$.

   $\diamond$    In $1+1$ dimensions, negative energy fluxes are allowed by energy conditions \cite{Ford_1995, Ford_1996, Ford_1999,Ford_2002, Fewster_Lectures_QEIs, Kontou_2020}, and the Fulling-Davies formula predicts them in the non-adiabatic regime where $\dot{k} < -\frac{1}{2}k^2$, \cite{Walker1985,FordRoman_MM_Neg_Energy}. While these effects are discarded in \cite{Carlitz-Wiley1987_Lifetime, Carlitz-Wiley1987_Reflections, Good_2013}, their implications for unitarity and black hole purification are discussed, e.g., in \cite{Bianchi:2014vea, Bianchi:2014qua, Walker1985, Abdolrahimi_2015, Parker, FordRoman_MM_Neg_Energy, Good_2015, Ford_2004,Agullo:2026,Chen_2017}. In particular, in \cite{Bianchi:2014qua,Bianchi:2014vea}, using the expression \eqref{eq:S-ent} for the entanglement entropy of Hawking radiation \cite{Bianchi:2014qua}, a puzzle was identified: Unitarity and the balance law implies that, before purification, a burst of negative energy density is necessarily emitted. This phenomenon results in a puzzling ``last gasp'' of an evaporating black hole which would increase its mass before disappearing. Remarkably, the correct identification of the Bondi mass and the radiative flux \eqref{eq:F-rad-quantum} resolve this puzzle: the flux is positive and the mass decreases monotonically \eqref{eq:M-dot-quantum} so that the black hole does in fact evaporate.

   $\diamond$ The expression \eqref{eq:F-rad-quantum} for the radiative energy flux in Hawking radiation was derived here by first identifying the radiative flux in the full $3+1$ dimensional classical theory  \eqref{eq:F-rad-mat}, and then using the correlation functions \eqref{eq:DxDx}--\eqref{eq:xDDx} at $\mathcal{I}^+$ computed in the mirror model for $3+1$ black hole evaporation. The energy flux $\langle F^{(\mathrm{rad})}_{\mathrm{mat}}\rangle(u)$ turns out to be manifestly positive, although this was not a priori required. It is interesting to compare this result to analogous ones discussed in the context of $2$d dilaton gravity, such as the Callan~et~al.~(CGHS) model \cite{Callan_1992} and the Russo~et~al.~(RST) model \cite{Russo:1992ht}. Both models present negative energy fluxes near the end of dilatonic-black-hole evaporation, which result in pathological behavior for its mass (see, for instance, \cite{TadaUeharaConsequencesHawking1995,KimLeeAdmMass1996, KimLeeHawkingRadiation1995} for CGHS and \cite{BilalKoganHamiltonianApproach1993,deAlwisTwodimensionalQuantum1994,Muller-KirstenEtAl2DQuantum1995} for RST). Ashtekar, Taveras, and Varadarajan (ATV) proposed a new expression $F_{\mathrm{ATV}}(u)$ for the flux of a CGHS dilatonic black hole which results in a quantum-corrected Bondi mass without any pathologies \cite{Ashtekar:2008jd,Ashtekar:2010hx,Ashtekar:2010qz}. The expression of the ATV mass was also derived in \cite{Barenboim:2025fds} through a Hamiltonian analysis of the semi-classical effective action for dilaton gravity with the Polyakov term \cite{PolyakovQuantumGeometry1981}.  Recently, in \cite{Varadarajan-2025}, Varadarajan advocated  the use of the ATV prescription for the energy flux computed in the mirror model for $3+1$ black hole evaporation with spherical symmetry. In this case, the ATV prescription can be expressed in terms of the peeling function \eqref{eq:ku}, $F_{\mathrm{ATV}}(u) =\frac{\hbar}{48\pi}k^2(u)=\langle F^{(\mathrm{rad})}_{\mathrm{mat}}\rangle(u)$ and, remarkably, it turns out to coincide with the renormalized expectation value of the radiative flux presented here.

\smallskip

\comment{The results presented in this Letter can be extended in various directions, which we discuss briefly here and in the companion paper \cite{BianchiParaizo2}:}

   $\circ$ We focused on the quantization of the flux of a massless minimally-coupled scalar field at $\mathcal{I}^+$. For such an asymptotic analysis, the radiative flux cannot register the Hawking radiation of massive particles, and a massive scalar field can only contribute to the renormalized expectation value of the Bondi mass of an evaporating black hole, not to its radiative flux. On the other hand, a conformally coupled scalar field would contribute only to the radiative flux and not to the mass. \comment{This is perhaps unsurprising given that %our prescription extracted the conformally-invariant part of the stress tensor flux Eq.~\eqref{eq:Fmatt}, but 
   the stress-tensor flux of a conformally-coupled field is already conformally invariant. Therefore, its classical and quantum balance laws would be given by equations \eqref{eq:M-dot}--\eqref{eq:M-Bondi} and \eqref{eq:M-dot-quantum}--\eqref{eq:M-quantum}, respectively, \textit{without} the additional matter-contribution to the charge.}

$\circ$  Crucially, the phenomenologically relevant contributions to evaporation are the radiative fluxes of electromagnetic and gravitational waves. It is important to repeat the analysis presented here, by going beyond the $s$-wave approximation and including the back-scattering potential. Specifically, a black hole of mass $M\approx 10^{12}\;\mathrm{kg}$ has a temperature of about $9\;\mathrm{MeV}$, and the power emitted in Hawking radiation in a neighborhood of the black hole was estimated by Page \cite{Page:1976df,Page:2004xp} to be $P\approx 6\times 10^9\;\mathrm{W}$, of which $45\%$ is in electrons and positrons, $45\%$ is in neutrinos $\nu_e\;\bar{\nu}_e$, $9\%$ is in photons, and $1\%$ is in gravitons. On the other hand, as electrons, positrons, and neutrinos are massive, the luminosity $\langle F^{(\mathrm{rad})}_{\mathrm{mat}}+F^{(\mathrm{rad})}_{\mathrm{grav}}\rangle$ measured far away---ideally at future null infinity---is only $1/10$ as large and consists approximately of $90\%$ photons and $10\%$ gravitons. Moreover, here we focused only on energy fluxes for the quantum theory, but the classical expressions \eqref{eq:F-Rad} and \eqref{eq:Q-phi} also apply to other BMS charges such as angular momentum and supertranslations \cite{Bondi:1962px,Flanagan:2015pxa, Compere:2019gft, CompereNotes}, which are expected to be relevant for infrared effects in black hole evaporation \cite{Hawking:2016msc,Flanagan_2021}.

   $\circ$ In the model considered here, we find that the Bondi mass includes an intriguing entropic contribution  $\tfrac{\hbar}{2\pi}\dot{S}_{\mathrm{ent}}^{(\mathrm{rad})}(u)$ given by  the entanglement entropy of scalar-field radiation, Eq.~\eqref{eq:M-entropic}. It is tempting to conjecture that this correction is universal. It would be interesting to investigate if a similar entropic contribution appears when one considers the Hawking flux of electromagnetic or  gravitational waves. As a simple test of the conjecture, one can consider  $N$ scalar fields, which results in a flux $\langle F^{(\mathrm{rad})}_{\mathrm{mat}}\rangle(u)=N\tfrac{\hbar}{48\pi}k^2(u)$ and an entanglement entropy $S_{\mathrm{ent}}^{(\mathrm{rad})}(u)=N\int_{-\infty}^u \frac{1}{12} k(u')\,du'$; thus, it turns out that the number of fields $N$ cancels in the entropic correction \eqref{eq:M-entropic}. \comment{Furthermore, this entropic correction bares a striking resemblance to the asymptotic Quantum Null Energy Condition \cite{Bousso:2016vlt}, a relation that would be interesting to investigate further.}

\smallskip

   $\circ$ The consequences of the balance law \eqref{eq:M-dot-quantum}--\eqref{eq:M-entropic} for the end-point of black hole evaporation and the fate of information are discussed in a companion paper \cite{BianchiParaizo2}, \comment{in which we analyze the Page curve for the entanglement entropy of radiation, and derive lifetime bounds}. We briefly point out here that the condition that Hawking radiation eventually stops, $\langle F^{(\mathrm{rad})}_{\mathrm{mat}}\rangle=0$, now implies that the ray-tracing function has the linear form $p(u) = A\,u+B$.  This condition is simpler than the modular form $p(u) = (Au+B)/(Cu+D)$ that one would obtain by demanding that the Fulling-Davies flux $F_{\mathrm{FD}}(u)$ vanishes. Furthermore, it implies that the geometry returns to flat spacetime after evaporation. These simple considerations fix the global causal structure following evaporation, are in-line with the arguments presented in \cite{Varadarajan-2025} and \cite{Bianchi:2018mml}, and are directly relevant to the purification of partner modes in the moving-mirror models \cite{Hotta_PartnerParticles,Tomitsuka_2020,Agullo_2024,Agullo:2026}. \comment{Furthermore, the flux \eqref{eq:F-rad-quantum} and entropy \eqref{eq:S-ent} immediately imply a minimum lifetime bound of $M^4/\hbar^{3/2}$ \cite{BianchiParaizo2}.}

%%%%

\bigskip

\noindent \emph{Acknowledgments}---We thank Abhay Ashtekar for valuable discussions on the distinction between radiative and Coulombic information; Matthew Brandsema and Kenneth Czuprynski for helpful comments on the manuscript and insightful discussions; \comment{and Robert Wald for discussions on his prescription with Zoupas}. We also thank an anonymous \comment{referee for pointing out the similarity of our results in the quantum theory with the Quantum Null Energy Condition applied to null infinity.} D.E.P. acknowledges support via the Bunton-Waller award from Penn State and the Walker fellowship from the Applied Research Lab. E.B. is supported by the National Science Foundation, Grants No. PHY-2513194. This work was made possible thanks to the support of the WOST project (\href{https://withoutspacetime.org}{\mbox{withoutspacetime.org}}), funded by the John Templeton Foundation (JTF) under Grant ID 63683.

%%%%

\bibliography{references}

\end{document}